\documentclass[fleqn,usenatbib]{mnras}

\usepackage{newtxtext,newtxmath}
\usepackage{cleveref}

\usepackage[T1]{fontenc}

\DeclareRobustCommand{\VAN}[3]{#2}
\let\VANthebibliography\thebibliography
\def\thebibliography{\DeclareRobustCommand{\VAN}[3]{##3}\VANthebibliography}

\usepackage{soul}
\usepackage[dvipsnames]{xcolor}

\usepackage{graphicx}	
\usepackage{amsmath}	
\usepackage{xcolor}		
\usepackage{physics}	
\usepackage{multicol}   
\usepackage{nicefrac}
\usepackage{csquotes}
\usepackage{threeparttable}

\title[Towards astrophysical 3rd order correlations]{Towards measuring astrophysical third order correlation functions with the H.E.S.S. optical intensity interferometer}

\author[A. Zmija et al.]{
Andreas Zmija$^{1,2}$\thanks{E-mail: andreas.zmija@univ-cotedazur.fr},
Gisela Anton$^{1}$,
Christopher Ingenhuett$^{1}$,
Alison Mitchell$^{1}$,
Prasenjit Saha$^{3}$,
\newauthor Pedro Silva Batista$^{1}$,
Naomi Vogel$^{1}$,
Adrian Zink$^{1}$,
Robin Kaiser$^{2}$,
Stefan Funk$^{1}$
\\
$^{1}$Erlangen Centre for Astroparticle Physics, Friedrich-Alexander-Universität Erlangen-Nürnberg, Nikolaus-Fiebiger-Str. 2, 91058 Erlangen, Germany\\
$^{2}$Institute de Physique de Nice, Université Côte d'Azur, 17 rue Julien Lauprêtre, 06200 Nice, France\\
$^{3}$Universität Zürich, Physik-Institut, Winterthurerstrasse 190, Zürich, Switzerland
}

\date{This is a pre-copyedited, author-produced PDF of an article accepted for publication in Monthly Notices of the Royal Astronomical Society following peer review. The article can be found at \url{https://doi.org/10.1093/mnras/staf2184}}

\pubyear{\the\year{2025}}

\begin{document}
\label{firstpage}
\pagerange{\pageref{firstpage}--\pageref{lastpage}}
\maketitle

\begin{abstract}
The closure phase, the sum of the three Fourier phases in a telescope triangle, is an important tool in astronomical interferometry, helping to reconstruct the geometries of the observed objects. While already established in amplitude interferometry, for the recently expanding field of intensity interferometers the closure phase enables recovering information of the interferometric phases that are otherwise inaccessible with this technique. To extract the (cosine of) the closure phase ($\cos \phi$) in intensity interferometry, third-order correlations between three simultaneously measuring telescopes have to be computed. As the signal-to-noise of such three-photon correlations is too small for current generation intensity interferometers, no third-order correlations of astrophysical targets have been recorded so far.
In this paper we present the first measurements of third order correlation functions of two stellar systems, Nunki ($\sigma $ Sgr) and Dschubba ($\delta$ Sco), observed with the H.E.S.S. intensity interferometer in 2023. We show how to isolate the three-photon contribution term from the two-photon contributions, in order to access $\cos \phi$. For the observed stellar targets the sensitivity is not high enough to extract closure phase information. To demonstrate that the analysis works well we further extract $\cos \phi$ in a laboratory experiment, using the H.E.S.S.\ intensity interferometer and a pseudo-thermal light source.
\end{abstract}

\begin{keywords}
instrumentation: high angular resolution -- instrumentation: interferometers -- techniques: interferometric -- stars: imaging -- methods: data analysis -- telescopes
\end{keywords}



\section{Introduction}

Within the last decade, the revival of stellar intensity interferometry -- anticipated by multiple astronomers \citep{lebohec2008toward, dravins2016intensity, rou2013monte} -- has gained a lot of momentum, supported by novel observational measurements. While the only successful demonstration of this technique prior to the new era dates back to the pioneering work of Twiss and \cite{brown1974intensity} between the late 1950s and the early 1970s, nowadays multiple telescope and telescopic arrays are performing stellar intensity interferometry observations, and have proven that using modern detectors, intensity interferometers complement the well-established amplitude interferometers in the field of high angular resolution measurements. These new intensity interferometers not only include the three Imaging Atmospheric Cherenkov Telescopes (IACTs) VERITAS \citep{abeysekara2020demonstration}, MAGIC \citep{acciari2020optical} and H.E.S.S.\ \citep{zmija2024first}, but also optical telescopes, such as the C2PU facility at the Calern Plateau \citep{lai2018intensity}, and even with two auxiliary telescopes of the VLTI \citep{matthews2022intensity}.

Intensity interferometry is a robust method for measuring stellar diameters,that is almost insensitive to atmospheric turbulence. Hence it can in principle be used at arbitrarily large interferometric baselines (multiple kilometers are often envisioned), outperforming amplitude interferometers in this regard (CHARA currently operates at the longest baseline of $330\,$meters \citep{ten2005first}, while NPOI plans on operating at a baseline of $432\,$meters \citep{baines2017fundamental}). On the other hand, the signal-to-noise is much lower than in amplitude interferometry, often requiring long exposures on targets as well as either large light collectors or very high time resolution.

The programs of modern intensity interferometers have already evolved beyond the measurement of one-dimensional uniform-disk angular diameters: while the extraction of limb darkened stellar diameters (albeit using model atmospheres) has become the standard reconstruction technique, some observatories have started analysing rotationally non-symmetrical objects. Both VERITAS \citep{rose2025first} and MAGIC (in combination with the Large Size Telescope 1 of the CTAO-North) \citep{garcia2025observation} observed the rapid rotator $\gamma$ Cas, extracting the oblateness of the star and its position angle.

A frequent criticism of intensity interferometry is its inability to measure the interferometric phase. This is less of a problem for observing simple objects like rotationally symmetric single stars, where the stellar geometries are non-ambiguously linked to the squared visibilities that intensity interferometers can measure, but becomes progressively disadvantageous as the observed objects contain a high amount of complexity in their intensity profiles. Missing the phase information prevents a clear reconstruction of the object's intensity distribution from the interferometric data, but only allows parameter estimations using a set of assumptions about the observed object. However, in arrays of at least three telescopes that record the light intensities simultaneously, the cosine of closure phase $\phi$, i.e. the sum of the interferometric phases in a triangle of telescopes, can be recovered. 
This allows for the recovery of at least parts of the phase information. The science case of fast-rotating stars in the context of intensity interferometry with closure phases was investigated by \cite{2015MNRAS.453.1999N}. They demonstrate that closure phase information benefits the image reconstruction.

More complicated star models, e.g. stars with star spots, or faint companions \citep{le2012sensitivity} rely on the availability of closure phases in the interferometric data. Utilising third-order intensity correlations for the measurement of the closure phase is therefore an important milestone in intensity interferometry. Due to the signal of three-photon correlations being even smaller compared to two-photon correlations, and the strongly increased computational effort, these measurements have not yet been performed by any of the astronomical intensity interferometers. Instead they are directed towards the (near) future, where large upcoming telescope arrays such as the CTAO promise to boost signal-to-noise, if equipped for intensity interferometry \citep{dravins2015long}.

In this paper, we present an astrophysical third-order correlation measurement, performed with the H.E.S.S.\ intensity interferometer during a measurement campaign in 2023. We used three of the $12\,$m diameter telescopes to compute the third-order correlation of two observed stellar systems, Nunki ($\sigma$ Sgr) and Dschubba ($\delta$ Sco). While the sensitivity of the data do not allow constraints on values for the closure phases, we show a proof of concept of an astrophysical third-order correlation measurement, introducing an analysis algorithm, able to calculate $\cos \phi$ in future high signal-to-noise measurements.

We complement the astrophysical measurements by a laboratory experiment, measuring the closure phase of a pseudo-thermal light source using the exact same interferometer that was used at H.E.S.S., proving that the interferometer and the algorithm are indeed able to extract $\cos \phi$ from the correlation measurement.

\section{Basics of first- to third- order correlations}

In an ideal amplitude interferometer, the first-order coherence function is calculated between two observers $i$ and $j$:

\begin{equation}
    g^{(1)}_{ij}(\tau,\mathbfit{b}) = \frac{\expval{E^*(t,\mathbfit{r}_i)E(t+\tau, \mathbfit{r}_i+\mathbfit{b})}}{\expval{E^*(t,\mathbfit{r}_i)E(t,\mathbfit{r}_i+\mathbfit{b})}},
\end{equation}

where $\tau$ is a time delay, and $\mathbfit{b} = \mathbfit{r}_j - \mathbfit{r}_i$ is the \textit{baseline vector} between the observers at positions $\mathbfit{r}_i$ and $\mathbfit{r}_j$. In this representation, the vectors $\mathbfit{r}_i$, $\mathbfit{r}_j$ and $\mathbfit{b}$ are projections from the line of sight of the observed target. 

At zero time-delay, the first order coherence function can by identified to be the complex visibility $\gamma_{ij}$. The polar representation

\begin{equation}
    g^{(1)}_{ij} (\tau=0) = \gamma_{ij} = A_{ij} e^{{\textnormal{i}\Phi_{ij}}}
\end{equation}

can be used to reconstruct the source geometry. According to \cite{van1934wahrscheinliche} and \cite{zernike1938concept}, the source intensity profile is connected to the first-order coherence function via a Fourier transform. Therefore we call $A_{ij}$ the Fourier amplitude, and $\Phi_{ij}$ the Fourier phase. In interferometry, these quantities are often referred to as visibility amplitude and visibility phase.

Intensity interferometers, in contrast, measure the second-order coherence function:

\begin{equation}
    g^{(2)}_{ij}(\tau,\mathbfit{b}) = \frac{\expval{I(t,\mathbfit{r}_i)I(t+\tau, \mathbfit{r}_i+\mathbfit{b})}}{\expval{I(t,\mathbfit{r}_i)}\expval{I(t,\mathbfit{r}_i+\mathbfit{b})}},
\end{equation}

with $I_{i,j}$ being the light intensities at observers $i$ and $j$. For thermal light sources, first- and second-order coherence functions are linked via the Siegert relation \citep{ferreira2020connecting}:

\begin{equation}
    g^{(2)} = 1 + |g^{(1)}|^2.
\end{equation}

For image reconstruction, it becomes evident that the utility of $g^{(2)}_{ij} (\tau=0) = 1 + |\gamma_{ij}|^2 = 1 + A^2_{ij}$ is limited, as only the (square of the) Fourier amplitude can be measured, but the Fourier phase is lost. This is a direct consequence of measuring intensities instead of electromagnetic waves.

The possibility to retrieve phase information arises in three-photon correlations from a triangle of telescopes $i,j,k$. This is the connected to the third-order coherence function:

\begin{align} \label{eq:g3}
    \begin{split}
    g^{(3)}_{ijk}(\tau_1, \tau_2, \mathbfit{b}_1, \mathbfit{b}_2) &= \frac{\expval{I(t,\mathbfit{r}_i)I(t+\tau_1, \mathbfit{r}_i+\mathbfit{b}_1)I(t+\tau_2,\mathbfit{r}_i+\mathbfit{b}_2)}}{\expval{I(t,\mathbfit{r}_i)}\expval{I(t, \mathbfit{r}_i+\mathbfit{b}_1)}\expval{I(t,\mathbfit{r}_i+\mathbfit{b}_2)}},
    \end{split}
\end{align}

with $\mathbfit{b}_1 = \mathbfit{r}_j-\mathbfit{r}_i$ and $\mathbfit{b}_2 = \mathbfit{r}_k-\mathbfit{r}_i$. According to \cite{malvimat2014intensity} we can express the third-order correlations via 1st and 2nd order correlations:

\begin{equation}
    g^{(3)}_{ijk} = 1 + {|g^{(1)}_{ij}|}^2 + {|g^{(1)}_{ik}|}^2 + {|g^{(1)}_{jk}|}^2 + 2 \textnormal{ Re} \left(g^{(1)}_{ij} g^{(1)}_{jk} g^{(1)}_{ki}\right).
\end{equation}

At zero-time delay ($\tau_1 = \tau_2 = 0$), $|g^{(1)}_{ij}|$ are the three Fourier amplitudes, and the real part of the triple-product of $g^{(1)}$ reveals the cosine of the closure phase $\phi = \phi_{ij} + \phi_{jk} + \phi_{ki}$:

\begin{equation}
    \label{eq:g3_amplitudes}
    A_{ijk} = 1 + A_{ij}^2 + A_{ik}^2 + A_{jk}^2 + 2 \cdot A_{ij} A_{ik} A_{jk} \cdot \cos \phi,
\end{equation}

In this representation, $A_{ij}^2$ are the squared visibilities, the main measuring quantity of a stellar intensity interferometer, and $A_{ij}$ are the `normal' visibilities, usually measured by an amplitude interferometer.

In order to access the cosine term of the closure phase, the product of the three visibility amplitudes has to be measured. The squared amplitudes are on the order of $\tau_c/\tau_e$, with $\tau_c$ being the coherence time of the light, and $\tau_e$ being the electronic time resolution. This fraction is usually on the order of $10^{-3}$ for narrow-band filtering in optical telescopes with excellent time resolution, to $10^{-6}$ or $10^{-7}$  IACTs, where narrow-band optical filtering is often not possible, and time resolution is limited by the anisochronicities of the telescope mirrors. The challenge of measuring these small amplitudes is even more pronounced when accessing the closure phase term, where the triple-product of the visibility amplitudes is on the order of $(\tau_c/\tau_e)^{3/2}$, which may be $10^{-9}$ to $10^{-11}$ at IACTs, and actually are of order $10^{-10}$ for H.E.S.S., as will be shown further below.

\section{Measurements with the H.E.S.S.\ intensity interferometer}

The H.E.S.S.\ intensity interferometer is an external camera, designed to fit onto the lid of the gamma-ray cameras of the 12 m diameter H.E.S.S.\ gamma-ray telescopes in Namibia. So far it was in operation during two measurement campaigns in 2022 and 2023 during times of bright moon -- during which normal gamma-ray observations cease -- accumulating a total of about $72.5$ hours of data on different stellar targets. The angular diameters of $5$ stars or star systems have been measured. While in 2022 only two of the Cherenkov telescopes (CT) were operated as an optical intensity interferometer (CT3 and CT4), in 2023 a third telescope (CT1) was included. The interferometer uses photomultipliers as detectors, whose photo-currents are continuously digitised every $1.6\,$nanoseconds and stored to disk for offline correlation. For details about the setup and the second-order correlation analysis see \cite{zmija2024first} and \cite{vogel2025simultaneous}.

\begin{figure}
    \centering
    \includegraphics[width=0.49\columnwidth]{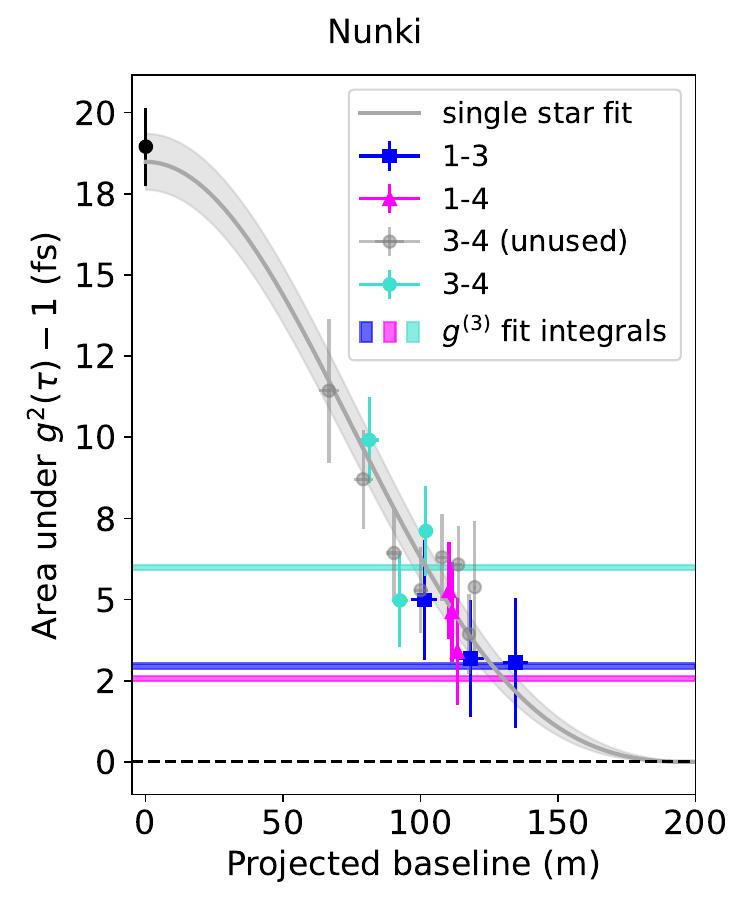}
    \includegraphics[width=0.49\columnwidth]{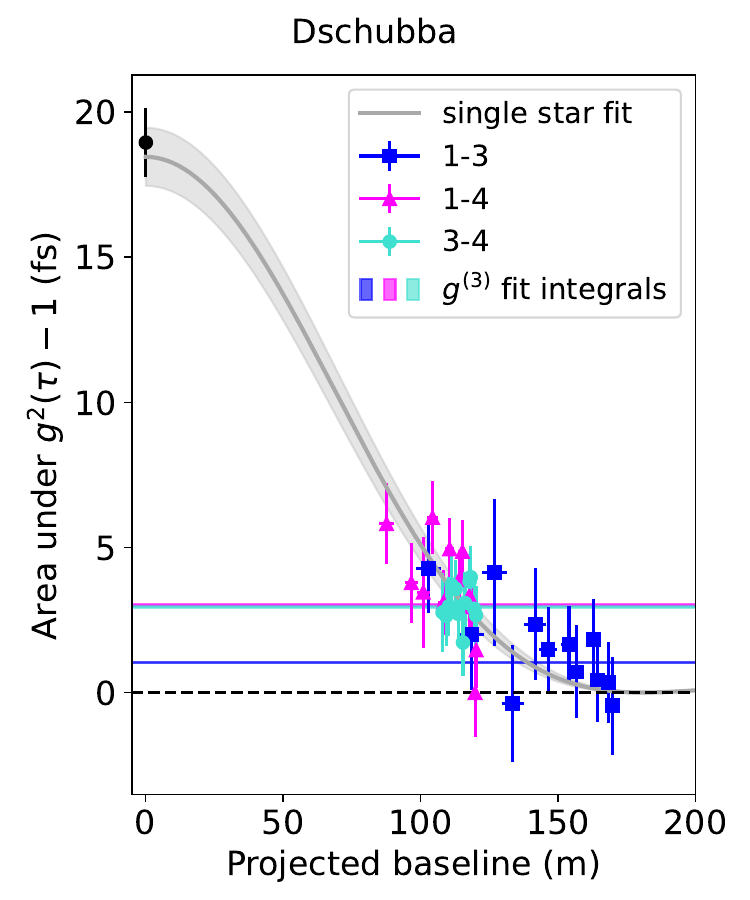}
    \caption{Second-order temporal correlation measurements of both stars at a wavelength of $470\,$nm. The full $g^{(2)}$ analysis of these datasets, including the determination of the data point at a projected baseline of $0$, can be found in \protect\cite{vogel2025simultaneous}.}
    \label{fig:g2}
\end{figure}

The storage of the raw data enables the possibility to not only calculate the second-order correlations that were used to extract the angular diameters of the systems, but also explore third order correlations, which are presented here. Two stellar systems were measured with three telescopes operating at the same time and are presented here: Nunki ($\sigma$ Sgr, mag$_B = 1.9$), where 105 minutes of three-telescope data were acquired, and Dschubba ($\delta$ Sco, mag$_B=2.2$), with an amount of 372 minutes of three-telescope data. In the case of Nunki, the total observation time in 2023 was actually 380 minutes, however, due to a broken amplifier, for 275 minutes only two telescopes at a time could be used, and hence for this time no three-photon correlations can be computed. The corresponding spatial second-order correlation measurements are displayed in Fig. \ref{fig:g2}, where the Nunki measurements contain data greyed-out, as for them only two telescopes were in operation.

The observation wavelength for both targets were $\lambda_0 = 470\,$nm, with a full-width at half maximum bandwidth of $\Delta\lambda = 10\,$nm. As described by \cite{vogel2025simultaneous}, a second wavelength band was also used at the same time ($\lambda_0=375\,$nm, $\Delta\lambda = 10\,$nm), but these data were not considered for a third order correlation analysis due to low signal-to-noise.

\section{results and analysis procedure}
\subsection{$g^{(3)}$ calculation}\label{sec:calculation}

\begin{figure}
    \centering
    \includegraphics[width=\columnwidth]{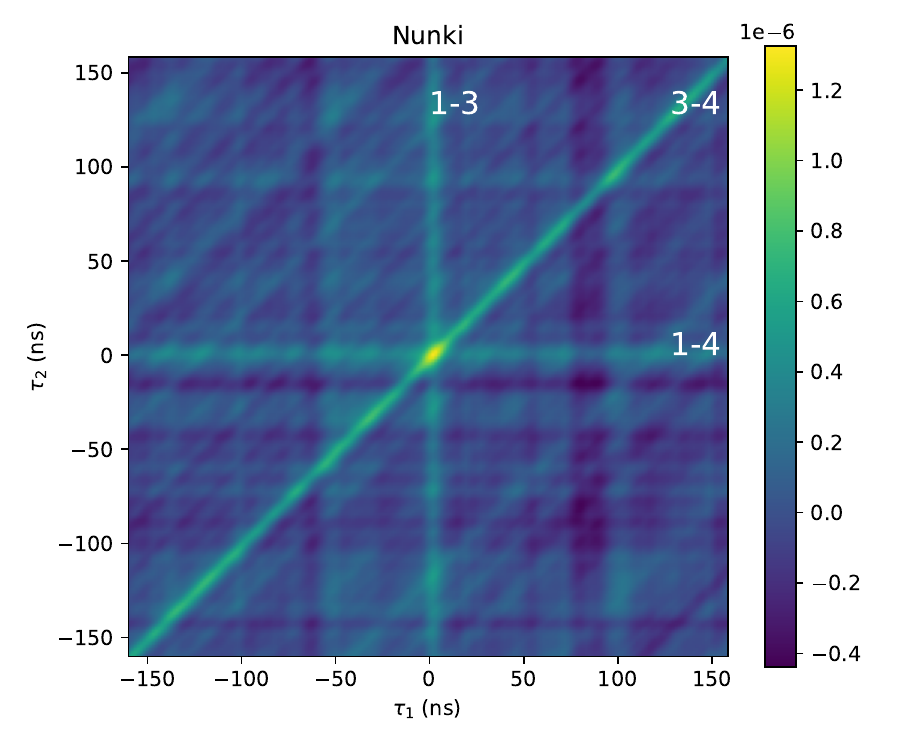}
    \includegraphics[width=\columnwidth]{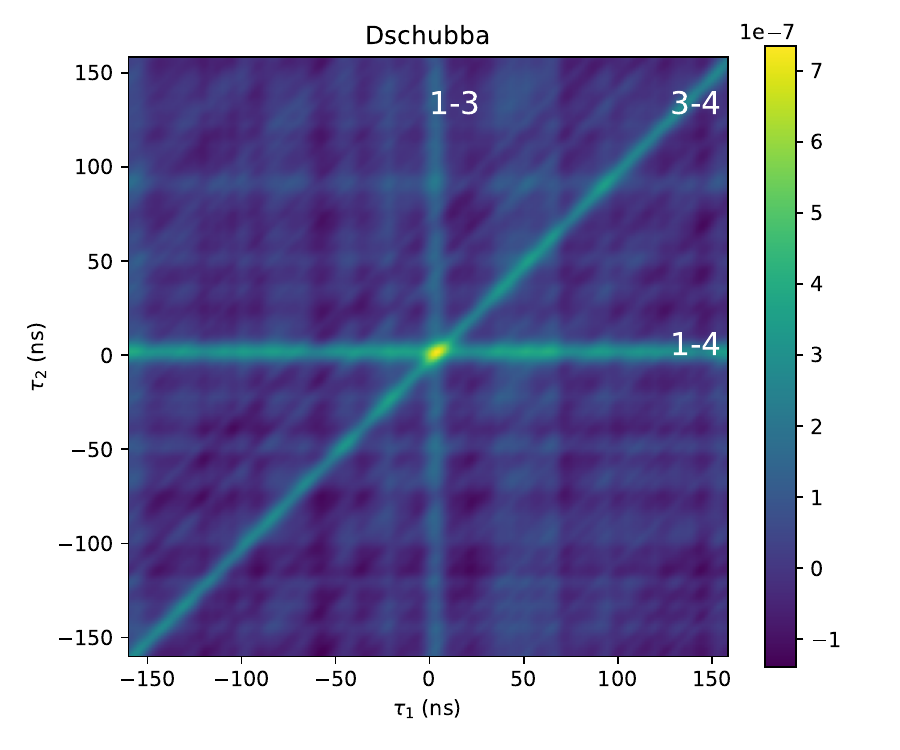}
    \caption{Third-order temporal correlation measurement of both stars at a wavelength of $470\,$nm. Both cases are integrated over the entire measurement time.}
    \label{fig:g3}
\end{figure}

Fig. \ref{fig:g3} shows the temporal third-order correlations of both targets. Following equation (\ref{eq:g3}), $g^{(3)}$ depends on two time delays. By our definition, $\tau_1$ is the time delay between CT1 and CT3, while $\tau_2$ is the time delay between CT1 and CT4. Consequently, the time delay between CT3 and CT4 is $\tau_2 - \tau_1$. Given that signal-to-noise is small for $g^{(3)}$ measurements, and to simplify the discussion of this demonstration, we chose not to split the data for the $g^{(3)}$ analysis into different time segments corresponding to different baselines, as we did for $g^{(2)}$, but instead calculate a single $g^{(3)}$ function per star, using all available three-telescope data.

The raw data consist of photo-currents $I_i(t_m)$, sampled from each photomultiplier ($i = 1,3,4$, according to the telescope number) every $\Delta t_m = 1.6\,$ns. We call these data streams \textit{waveforms}. Since the photon rates are on the order of GHz, and photon pulses spread over several time bins, multiple photon signatures are present within one time bin, and we cannot identify individual photons in the data. The waveforms are stored in binary files of $2\,$gigasamples size ($\approx 3.4\,$s length) per telescope. For these files the (unnormalised) three-photon correlation for any $\tau_1 \tau_2$ pair is computed by calculating the product of each waveform triple:

\begin{equation}
    G^{(3)} (\tau_1, \tau_2) = \sum_m I_1(t_m) \, I_3(t_m+\tau_1) \,  I_4(t_m+\tau_2)
\end{equation}

This calculation is a major computational effort. Although calculations of second-order correlations can be performed efficiently on GPUs by multiplying the Fourier transforms of the waveforms, such a trick is not possible for triple correlations. Additionally, the two-dimensional nature of the three-photon correlations increases the amount of time bins that have to be calculated by one dimension compared to two-photon correlations. We transferred the waveforms to the high-performance computation cluster at FAU Erlangen-Nürnberg, where the calculations are parallelised. Nevertheless, for the computation of third order correlations with a range of 200 time bins for each $\tau$, the computation time is roughly a factor $200$ times greater than the data acquisition time. This makes real-time third order correlations of photo-currents a major computational challenge. Future intensity interferometers may be based on time-tagging of single photons at smaller optical bandwidths, and FPGA based correlations, in order to decrease the computation time.

There exists an 8x8 bin periodic systematic pattern, originating from the ADC. This pattern is also present in one dimension and already well understood in the context of $g^{(2)}$. To eliminate it in $g^{(3)}$, we create an 8x8 bin template by averaging over 8x8 sections in a region of $G^{(3)}$ where no correlation signal is expected or visible. We use the bottom right corner of the plots presented in Fig. \ref{fig:g3}: $\tau_1 \in (19.2\,\textnormal{ns}, 160\,\textnormal{ns})$, and $\tau_2 \in (-160\,\textnormal{ns}, -19.2\,\textnormal{ns}$). Next we slide this template as a corrector over the entire $G^{(3)}$ map. This algorithm removes the 8x8 bin pattern, and further normalises the $G^{(3)}$ values. Finally, subtracting $1$ from each data point yields the $g^{(3)}-1$ data that are visible in Fig. \ref{fig:g3}. 

The plots display some characteristic features: three significant lines are visible, indicating photon correlations above the random level. The lines manifest around $\tau_1 = 0$ (vertical), $\tau_2 = 0$ (horizontal), and $\tau_1 = \tau_2$ (diagonal), and visualise the two-photon correlations between each pair of telescopes. In other words they display the $g^{(2)}$ amplitudes. The white labels on the lines point to the corresponding $g^{(2)}$ amplitude that is measured. Since the different telescope combinations have on average different projected baselines, the lines have different brightnesses, as apparent in Fig. \ref{fig:g2}. In both cases, the 1--3-correlation is comparatively faint due to its projected baseline being larger, while especially in the case of Nunki, the baseline 3--4 gives the highest correlation values due to the baseline being on average the smallest of the three. Quantitative values will be given in section \ref{sec:closure_phase_analysis}.

\subsection{Closure phase analysis}\label{sec:closure_phase_analysis}

The closure phase is -- in principle -- accessible via the central region at $\tau_1 = 0, \tau_2 = 0$. While the three previsously described lines resemble simultaneous two-photon correlations between pairs of telescopes, a simultaneous three-photon correlation gives rise to an additional term, increasing or decreasing the height of the correlation signal, depending on the value of the closure phase. This is reflected in the last term in equation (\ref{eq:g3_amplitudes}), where the cosine of the closure phase may be positive or negative. We established a default procedure of how to access the closure phase term by subtracting the other contributions of equation (\ref{eq:g3_amplitudes}). While the plots in Fig. \ref{fig:g3} already show $g^{(3)} - 1$, there are still the two-photon correlations to be removed.

A two dimensional fit is applied to the third-order correlation data, consisting of three `gaussian tubes' -- that is a Gaussian elongated along some axis. This is justified for the H.E.S.S.\ measurements, given that the timing resolution of the telescopes (optical time dispersion) convolved with the photon pulse correlations are gaussian shaped to very good approximation. As an example, the second order correlations between telescopes 1 and 4 are gaussian shaped around $\tau_2$, but independent of $\tau_1$, which leads to the tube-like structure. The entire fit function reads

\begin{align}
\begin{split}
    f(\tau_1, \tau_2) &= a_{13} \cdot \exp \left(-\frac{(\tau_1 - \mu_{13})^2}{2\sigma_{13}^2}\right)
                      + a_{14} \cdot \exp \left(-\frac{(\tau_2 - \mu_{14})^2}{2\sigma_{14}^2}\right)\\
                      &+ a_{34} \cdot \exp \left(-\frac{(\tau_1 - (\tau_2 + \mu_{14} - \mu_{13}))^2}{2\sigma_{34}^2}\right).
\end{split}
\end{align}

Small differences in the length of the signal cables in the different telescopes lead to slight offsets (time delays) of the center of the Gaussians from $\tau_1=0, \tau_2=0$ and $\tau_1 = \tau_2$, so that the center values $\mu_{13}$ and $\mu_{14}$ need to be included in the fit. The third time delay, the diagonal offset of the 3--4-correlation function from $\tau_1 = \tau_2$, is a direct consequence of the two other delays and is hence not a free fit parameter.

\begin{figure*}
    \centering
    \includegraphics[width=2\columnwidth]{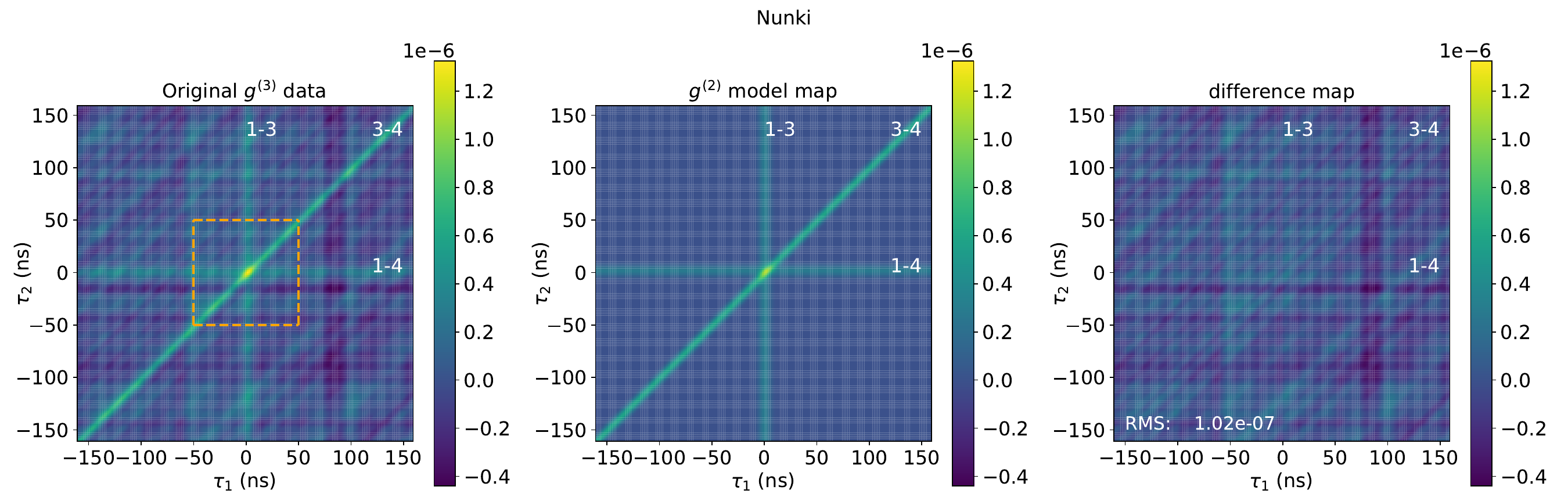}
    \caption{Analysis of the $g^{(3)}$ data for extracting the non-trivial three-photon contribution of Nunki. Left: original $g^{(3)}$ data. The dashed orange rectangle indicates the region that is excluded for the 2-dimensional fitting of the $g^{(2)}$ contributions. Center: The two-photon model, constructed from the results of the 2-dimensional fit of three gaussian functions along the $g^{(2)}$ axes. Right: The difference between the original data and the 2-d fits.}
    \label{fig:residuals_Nunki}
\end{figure*}

\begin{figure*}
    \centering
    \includegraphics[width=2\columnwidth]{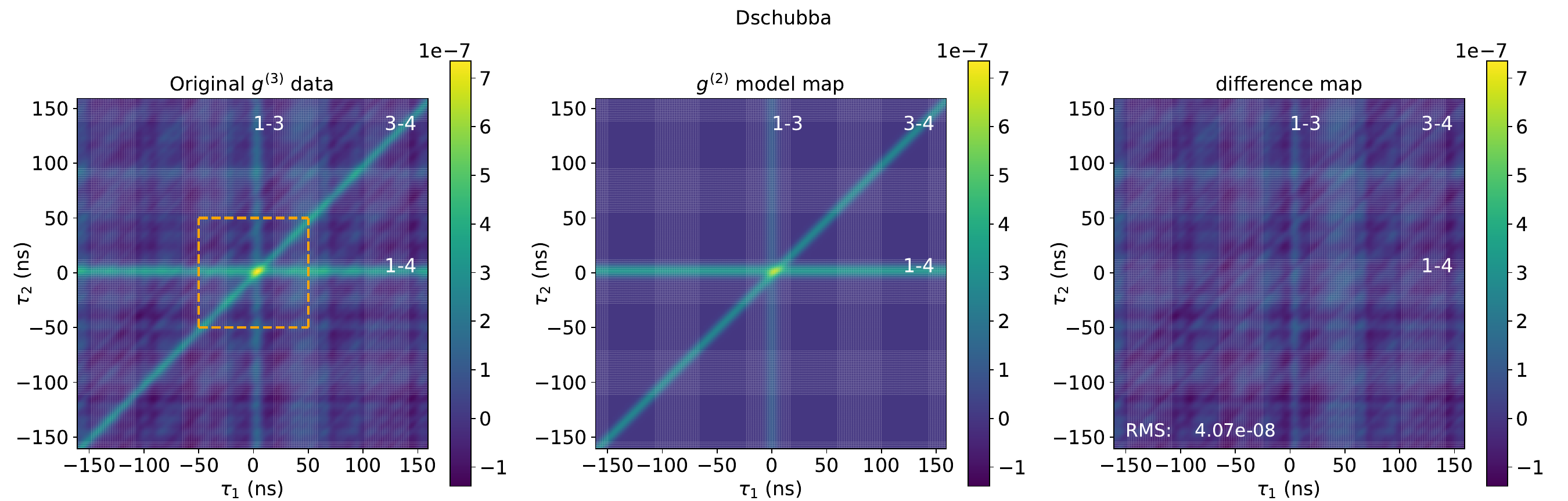}
    \caption{Analysis of the $g^{(3)}$ data for extracting the non-trivial three-photon contribution of Dschubba. Left: original $g^{(3)}$ data. The dashed orange rectangle indicates the region that is excluded for the 2-dimensional fitting of the $g^{(2)}$ contributions. Center: The two-photon model, constructed from the results of the 2-dimensional fit of three gaussian functions along the $g^{(2)}$ axes. Right: The difference between the original data and the 2-d fits.}
    \label{fig:residuals_Dschubba}
\end{figure*}

To exclude a potential effect of the three-photon correlations on the fit, the central region (a square with side length of $100\,$ns around the center) is excluded from the fit. This is visualised by the orange dashed square in the left panel in Figs. \ref{fig:residuals_Nunki} and \ref{fig:residuals_Dschubba}.

The two-dimensional fit characterises the two-photon correlations of all the telescope pairs. The results of the fit are displayed in the central plots in Figs. \ref{fig:residuals_Nunki} and \ref{fig:residuals_Dschubba} for Nunki and Dschubba respectively, the Gaussian fit amplitudes are listed in Table \ref{table:fit_amplitudes}.

Given that we set up a new algorithm for the calculation of $g^{(3)}$, we perform a consistency check by comparing the amplitudes, or the corresponding gaussian integrals, with the traditional $g^{(2)}$ measurements. We visualise the fit integrals from the $g^{(3)}$ fit, including their uncertainties, as horizontal bands in Fig. \ref{fig:g2}. These values coincide well with the average over the $g^{(2)}$ data points of each telescope combination.

The fit amplitudes of the Gaussians $a_{ij}$ can further be identified with the squared visibility amplitudes $A^2_{ij}$ from equation (\ref{eq:g3_amplitudes}). Subtracting these two-photon contributions from the $g^{(3)}$ data leaves us with the three-photon correlation in the centre of the difference map:

\begin{equation}
    A_{134} - 1 - A_{13}^2 - A_{14}^2 - A_{34}^2 = 2 \cdot A_{13}A_{14}A_{34} \cdot \cos \phi_{134}
\end{equation}

 The subtraction of the two-photon correlations from $g^{(3)}$ is shown in the right plots in Figs. \ref{fig:residuals_Nunki} and \ref{fig:residuals_Dschubba}. It can be seen that the fit procedure eliminates the two-telescope contributions, with no significant three-photon contribution present in the data. This latter finding can be easily motivated by estimations of the expected signal height: the strongest signals are expected either for closure phases of $\phi = 0$ and $\cos \phi = 1$, or for $\phi = -180^\circ$ and $\cos \phi = -1$. But even in these cases the necessary sensitivity in the difference map needs to be better than $2A_{13}A_{14}A_{34}$, a value that can be computed from the fitted squared amplitudes, and is also listed in Table \ref{table:fit_amplitudes}.

\begin{table}
\centering
\caption{Fitted squared visibility amplitudes for both stars, and expected maximal signal amplitude of the three-photon correlation term.}
\begin{tabular}{ c  c  c } 
 \hline
Amplitude & Nunki & Dschubba\\ [0.1ex] 
 \hline
 \noalign{\smallskip}
 $A_{13}^2$ &  $(2.95 \pm 0.05) \cdot 10^{-7}$ &  $(1.38 \pm 0.04) \cdot 10^{-7}$ \\
 $A_{14}^2$ &  $(2.79 \pm 0.05) \cdot 10^{-7}$ &  $(3.67 \pm 0.03) \cdot 10^{-7}$ \\
 $A_{34}^2$ &  $(5.93 \pm 0.05) \cdot 10^{-7}$ &  $(2.36 \pm 0.03) \cdot 10^{-7}$ \\
 \hline
 $2A_{13}A_{14}A_{34}$ & $(4.42 \pm 0.06) \cdot 10^{-10}$ & $(2.19 \pm 0.04) \cdot 10^{-10}$ \\
 \hline
\end{tabular}

\label{table:fit_amplitudes}
\end{table}

The necessary sensitivities for constraining the closure phase is in both cases on the order of $10^{-10}$, while the achieved sensitivities, expressed by the root mean square (RMS) of the difference maps, are $1.02 \cdot 10^{-7}$ for Nunki, and $4.07 \cdot 10^{-8}$ for Dschubba -- essentially 2-3 orders of magnitude worse. Hence, we are unable to provide a constraining estimation on the closure phase for these measurements. Quantitative discussion on necessary measurements times and way on how to improve signal-to-noise will be given in section \ref{sec:discussion}.

\section{Discussion of the closure phase measurements}

\begin{figure*}
    \centering
    \includegraphics[width=2\columnwidth]{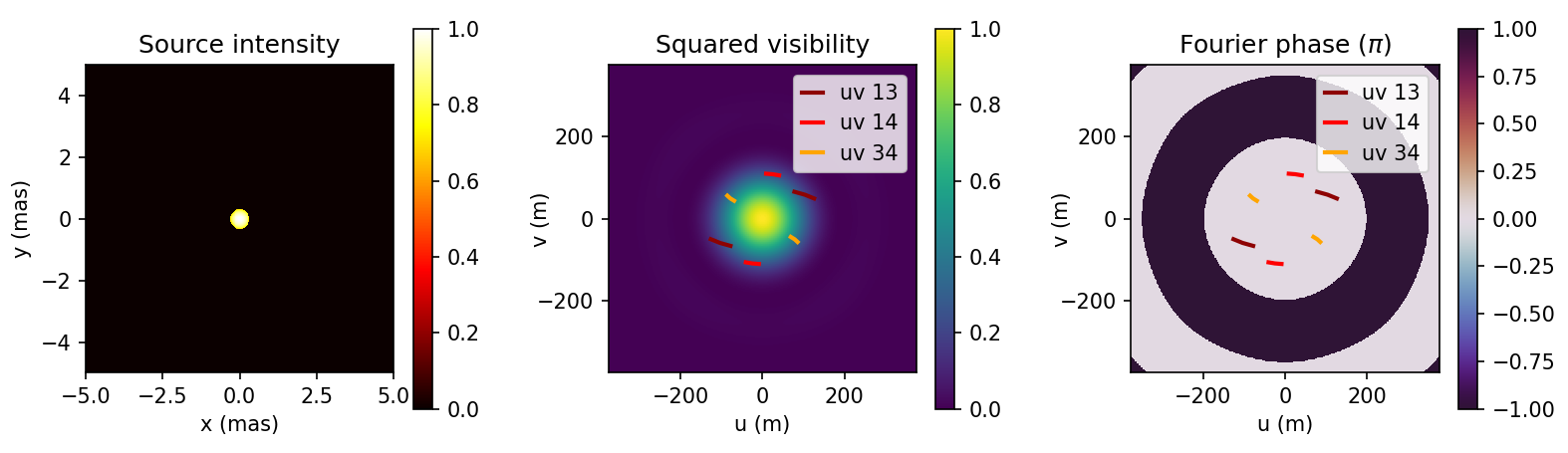}
    \caption{Reconstructed parameters for Nunki according to the single star limb-darkened model from \protect\cite{vogel2025simultaneous}. From left to right: the intensity profile of the star, squared visibility map on earth, Fourier phase map on earth. In the latter two maps, the uv tracks of the H.E.S.S.\ measurements are also shown}
    \label{fig:nunki_single}
\end{figure*}

\begin{figure*}
    \centering
    \includegraphics[width=2\columnwidth]{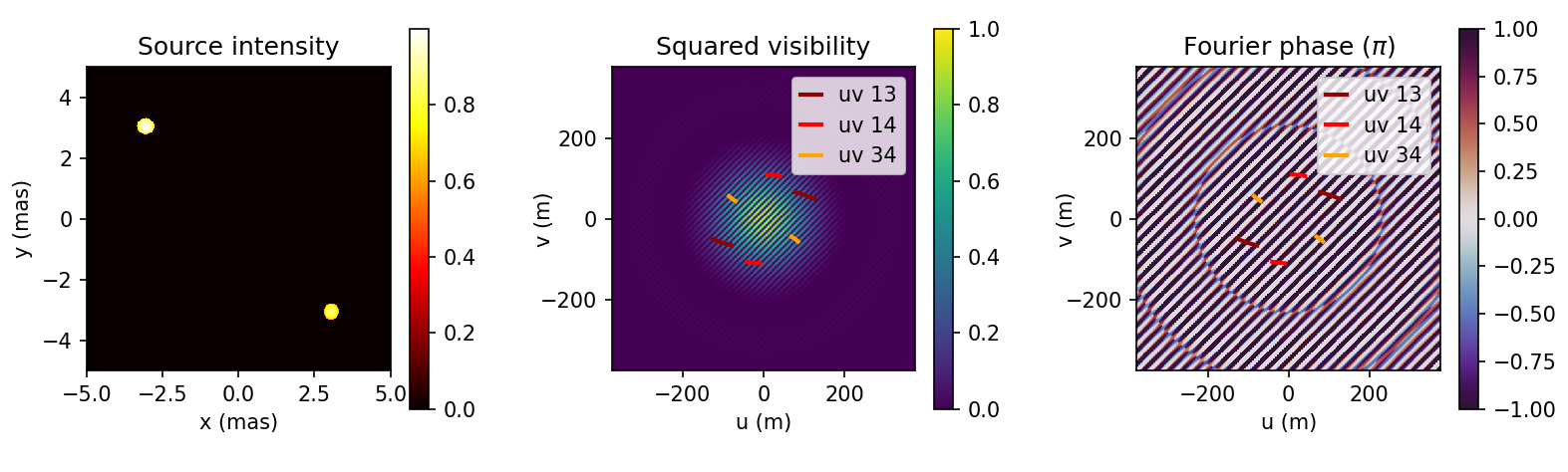}
    \caption{Reconstructed parameters for Nunki according to recent measurements of the VLTI \protect\cite{waisberg2025hidden}. From left to right: the intensity profile of the stars (the orientation is random), squared visibility map on earth, Fourier phase map on earth. In the latter two maps, the uv tracks of the H.E.S.S.\ measurements are also shown. The tracks are the same as in Fig. \ref{fig:nunki_single}.}
    \label{fig:nunki_binary}
\end{figure*}

Even though the sensitivities of the H.E.S.S.\ measurements do not allow for constraining closure phase information, we want to quickly discuss the value of such a measurement in a (yet futuristic) case where measurement parameters are tuned such that a measurement of $\cos \phi$ is actually possible. We take the measurements of Nunki as an example.

In \cite{zmija2024first} and \cite{vogel2025simultaneous} we modeled Nunki as a single star, as the squared visibility data indicated no particular contribution of multiple components. The measurements from 2023 indicated an angular diameter of $(0.63 \pm 0.02)\,$mas using a single star limb-darkened fit model. Recent measurements at the VLTI, however, indicate that Nunki is actually a binary, with both components being very similar in angular diameter ($0.55$ and $0.52\,$mas with an uncertainty of $0.07\,$mas, and a flux ratio of $0.88$ in the K band) separated by $8.6\,$mas \citep{waisberg2025hidden}. The binary nature of Nunki was already assumed by \cite{hanbury1974angular}, as the zero-baseline correlation value in their measurements was too small to be consistent with a single-star. This is not a major contradiction to the H.E.S.S.\ measurements, as the separation of the two components results in spatial oscillations of the correlation pattern that are so rapid that the H.E.S.S.\ telescopes with a dish diameter of $12\,$meters average over them, and are only sensitive to the envelope created by the angular size of the individual stars.

Two scenarios are modeled in two dimensions: in Fig. \ref{fig:nunki_single} for the single star case, and in Fig. \ref{fig:nunki_binary} for the binary case (a random orientation of the binary was chosen in this case). On the left, the intensity profile of the limb-darkened stars are shown. The other two images are created via numerical Fourier transforms from the intensity profile. The square of the Fourier amplitude (the squared visibility) is plotted in the center images, the Fourier phases, which cannot be directly measured with an intensity interferometer, are shown in the right images. The actual uv tracks of the three-telescope Nunki measurements are added to the squared visibility and fourier phase map.

\begin{figure}
    \centering
    \includegraphics[width=0.85\columnwidth]{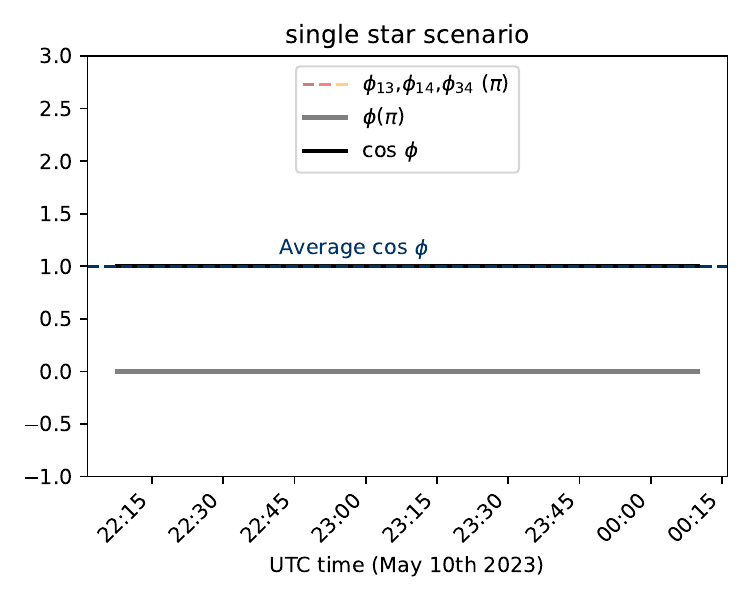}
    \includegraphics[width=0.85\columnwidth]{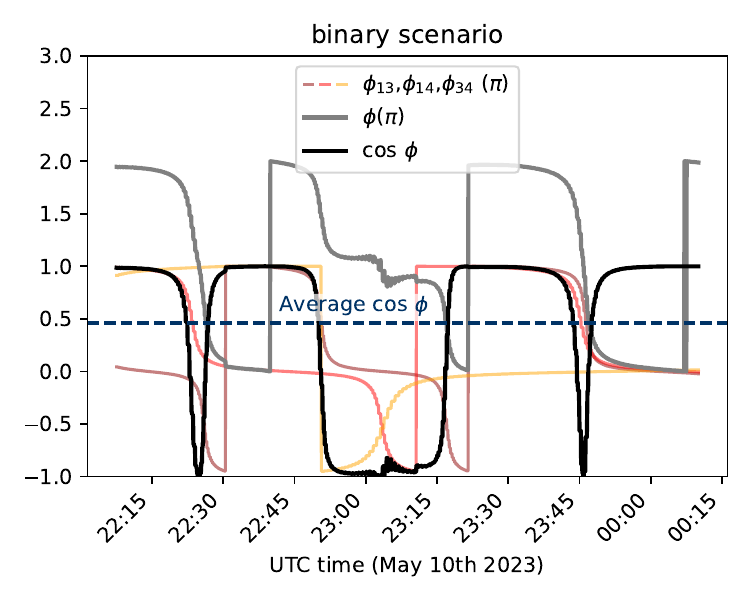}
    \caption{Temporal evolution of the closure phase for the scenario of a single star (left) and a binary with the orientation shown in Fig. \ref{fig:nunki_binary}.}
    \label{fig:closure_phase_scenarios}
\end{figure}

Unlike in a typical $g^{(2)}$ measurement, where the observation time is divided into multiple segments of different projected baselines/uv positions, for this $g^{(3)}$ calculation all available data were accumulated into a single data point. That means the closure phase measurement is also averaged over the total integration time, and so over all uv tracks. A time series of the three Fourier phases, the closure phase and its cosine is shown in Fig. \ref{fig:closure_phase_scenarios} for the single star scenario in the upper panel, and a binary scenario in the lower panel. This simplified model assumes infinitely small sizes of the telescopes, so that the uv tracks in Figs. \ref{fig:nunki_single} and \ref{fig:nunki_binary} have no lateral extension.

In the case of a single star, as depicted in figure, every part of the uv tracks of every telescope combination is clearly in the regime where each interferometric phase is zero, and so is the closure phase. That means the expected value of the closure phase is $\phi =0$ and $\cos \phi = 1$ for the single star model.

In case of the binary star, however, the telescopes clearly pass through areas of different Fourier phases, and they are expected to see jumps in the closure phase between $0$ and $180^\circ$. In a time-averaged measurement these jumps between $\cos \phi = 1$ and $\cos \phi = -1$ average to a value of $|\cos \phi| < 1$, in the assumed scenario it results in $\expval{\cos \phi} = 0.46$.

In reality, the $12\,$m dish size of the telescopes washes out the clear closure phase signals further, inducing also a spatial averaging over the sharp closure phase structures. In this way it is apparent how a determination of $\cos \phi$ can reveal differences between a single star and a binary.

Lastly, it should be mentioned that just like in other interferometric measurements it would be beneficial to divide the total measurement time into smaller segments, in order to obtain a finer binning of $g^{(2)}$ and $g^{(3)}$ in uv. This, however, comes at the cost of reduced signal-to-noise of each single measurement, an optimization problem that needs to be considered in the context of closure phase measurements in future intensity interferometers.

A similar discussion about Dschubba could be included here, however, due to the complexity of the stellar system, including a binary with very eccentric orbit, a circumstellar disc, and a potential third component \citep{miroshnichenko20132011}, we refrain from it in this paper.

\section{Closure phase measurements in the laboratory}

\begin{figure*}
    \centering
    \includegraphics[width=1.7\columnwidth]{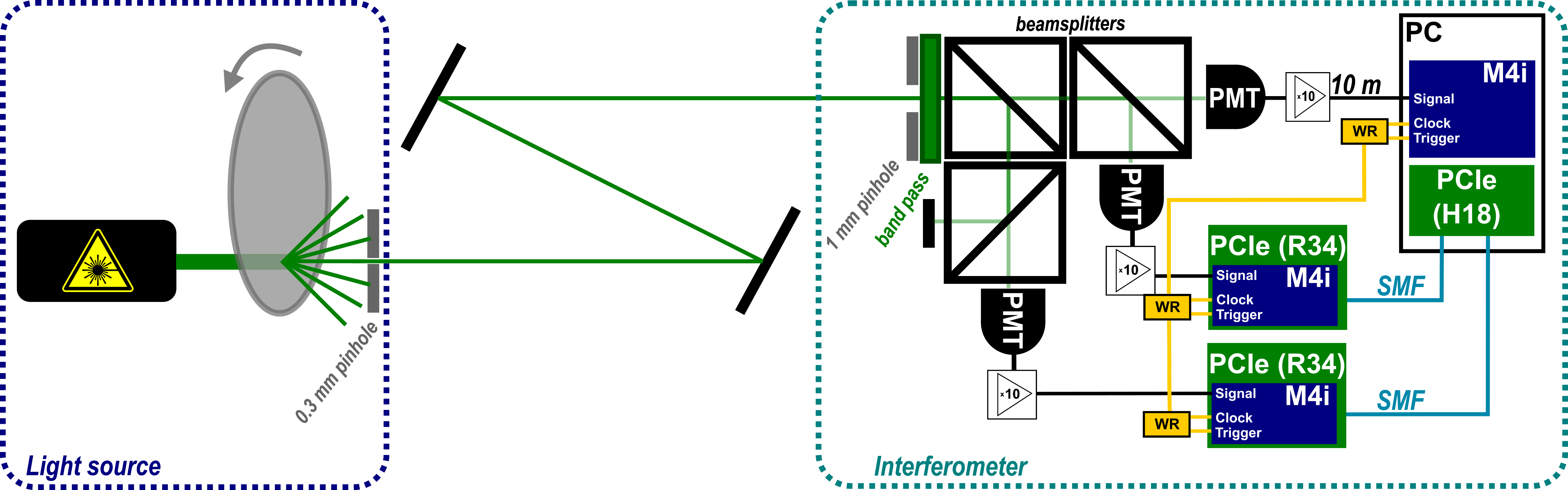}
    \caption{Laboratory setup to measure third-order correlations of the pseudo-thermal light source.}
    \label{fig:lab_setup}
\end{figure*}

In order to prove the H.E.S.S.\ intensity interferometer is capable of measuring a closure phase signal in the third order correlation, we brought the interferometer into the lab and performed $g^{(3)}$ measurements in a controlled environment. This was possible since the interferometer was dismounted from the telescopes after the campaigns and brought back to Erlangen in Germany. We reassembled the interferometer in the laboratory, using the same photo-multipliers, amplifiers and digitisers in a three-beamsplitter arrangement. The setup is sketched in Fig. \ref{fig:lab_setup}. We circumvent the need for extremely long exposures by the use of a pseudo-thermal light source with long coherence time. This is usually done by focusing a laser beam on a scattering medium, such as a collodial suspension \citep{carlile2021seeing}, or a rotating ground glass diffuser \citep{zhou2010third}. We use the latter, a disk attached to a rotational mount, where the coherence time can be set by the rotational speed of the disk, and is typically on the order of (tens of) microseconds. This leads to a much greater value of $\tau_c/\tau_e$, and therefore high signal-to-noise second- and third-order measurements already after seconds of data acquisition.

The light that is scattered by the disk is spatially filtered by a $0.3\,$mm diameter circular `exit' pinhole, in order to create a small, well defined light source. With the help of two mirrors, the light from the pinhole is guided to the beamsplitter setup, located at a distance of $5.5\,$meters. In front of the first beamsplitter we place an optical filter with central wavelength of $532\,$nm and optical bandwidth of $1\,$nm, in order to isolate the $532\,$nm laser light and prevent stray light. We also place a $1\,$mm diameter `entrance' pinhole in front of the first beamsplitter to restrict the effective size of the detectors. If the exit pinhole can be considered a perfect source of pseudo-thermal light, the squared visibility amplitudes would be expected to drop to 1 at a baseline of $11.9\,$mm, and hence the detectors stay within the first lobe of the Airy pattern due to the entrance pinhole. That means, the phases are expected to be $0$ over the detector area, and the closure phase is also expected to be $\phi = 0$ resulting in $\cos \phi =1$ and a maximum amplitude in $g^{(3)}$.

By trying different pinhole configurations we noticed that the assumption of the exit pinhole as a perfect source of pseudo-thermal light is not always justified, as we saw $g^{(2)}$ values differing from the expected ones. This may indicate that in some constellations the photon statistics of such an experiment is not identical with actual thermal light sources. Such deviations have been reported in similar experiments \citep{clark2024measuring}. In the constellation we present, we checked that the number of photons per time in each of the channels is distributed exponentially, and therefore assume a high similarity of the photon statistics with thermal light.

In contrast to the H.E.S.S.\ measurements, the number of photons hitting the detectors in the laboratory is very small ($\approx 1\,$MHz) due to the small pinholes. Instead of correlating the photo-currents, we decided to extract the arrival times of the photons from the waveforms by time tagging the peak positions of the individual photon pulses. We then perform correlations of the arrival times of the photons in order to calculate $g^{(2)}$ and $g^{(3)}$. In this way, we avoid implications by slight changes in the amplifier offsets, which can significantly affect the signal height in $g^{(2)}$ and $g^{(3)}$ in cases where the waveforms are not dominated by photon pulses. With this mechanism we also don't need to perform the 8x8 bit pattern correction, and normalization can be done mathematically by the product of the mean photon rates in the measurements.

\begin{figure*}
    \centering
    \includegraphics[width=2\columnwidth]{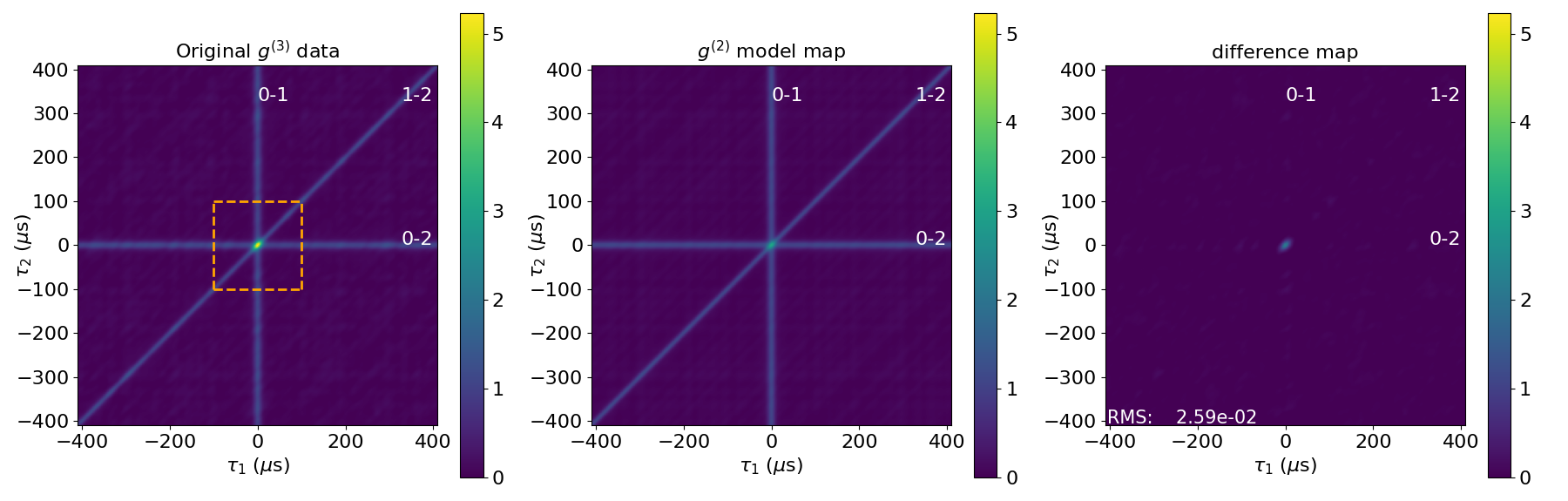}
    \caption{Laboratory measurement with the rotating ground glass disk. The center in the difference map reveals the non-trivial three-photon contribution.}
    \label{fig:residuals_lab}
\end{figure*}

Fig. \ref{fig:residuals_lab} shows the three-photon correlation maps as previously introduced. Due to the shape of the $g^{(2)}$ functions not being gaussian, and not easily described any analytic function, we take the interpolated shape resulting from the the calculation of $g^{(2)}$ as shape for the three tube fits in $g^{(3)}$. The comparison between the $g^{(3)}$ data in the left plot and the fit map in the center plot of the figure already indicates that the three $g^{(2)}$ amplitudes don't add up to the value of $g^{(3)} (0,0)$. This suspicion is confirmed in the difference map, where the three-photon correlation now shows a remaining contribution in the center, associated with the three photon correlation term $2 \cdot A_{13}A_{14}A_{34} \cdot \cos \phi_{134}$.

\begin{figure}
    \centering
    \includegraphics[width=0.8\columnwidth]{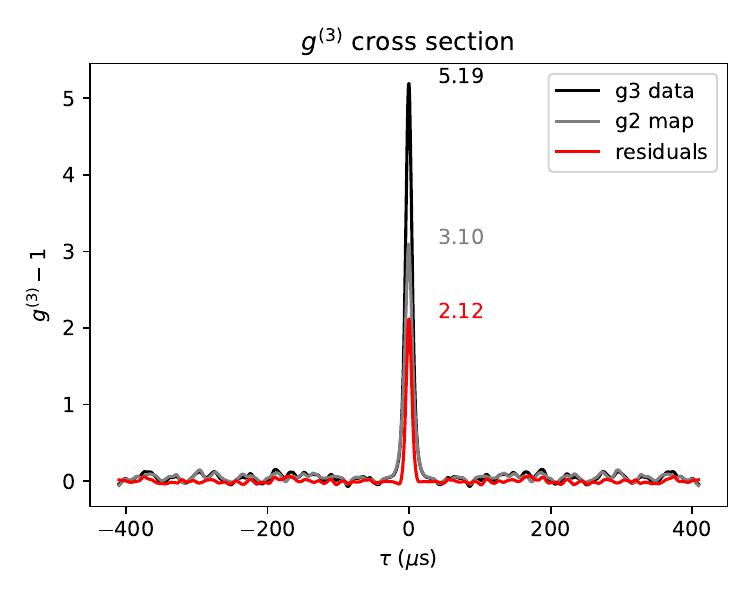}
    \caption{Cross section through the maps in Fig. \ref{fig:residuals_lab} from the top left to the bottom right.}
    \label{fig:residuals_lab_cross_section}
\end{figure}

A cross-section view from the top left corner to the bottom right corner (inspired by \cite{zhou2010third}) of each of the three maps is also shown in Fig. \ref{fig:residuals_lab_cross_section}.

From the data we can extract the fit squared visibility amplitudes $A_{ij}^2$ and the peak value of the third-order correlation data $P =\max (g^{(3)} - 1)$, we can calculate the cosine of the closure phase as

\begin{equation}
    \cos \phi = \frac{P}{2\sqrt{A_{01}^2 A_{02}^2 A_{12}^2}} = \frac{5.19}{2\sqrt{1.04 \cdot 1.04 \cdot 1.00}} = 1.02,
\end{equation}

which is in good approximation matched with the expected value of $\cos \phi = 1$ for $\phi=0$, as expected in the experiment.

\section{Discussion}\label{sec:discussion}

The laboratory experiment shows the applicability of the analysis algorithm to a three-photon correlation measurement. The H.E.S.S.\ telescope measurements are limited in their capability to measure $\cos \phi$ from actual thermal light due to its very small coherence time. For Nunki the sensitivity is $230$ times worse than the the order of magnitude of the closure phase signal, for Dschubba it is a factor of $186$. For a measurement of $\cos \phi$ to a precision of $0.1$, which we set as the goal for this discussion, the improvement of the sensitivity would need to be by a factor of $2300$ and $1860$ respectively.

There are several ways to improve the signal-to-noise ratio (S/N),
which depends on the observational setup as \citep[see e.g.,][]{2015MNRAS.453.1999N}
\begin{equation}
    \textnormal{S/N}_{g^{(3)}} \propto \frac{1}{\tau_e} \sqrt{\frac{A^3 T}{\Delta\lambda}},
\end{equation}
where $\tau_e$ is the interferometer's time resolution, $A$ is the telescope's collection area times the throughput, $T$ is the observation time, and $\Delta\lambda$ is the optical bandwidth.  In contrast to the two-photon case, the three-photon S/N depends on the optical bandwidth $\Delta\lambda$, and shorter bandwidths are preferred, up to the limit $\Delta\lambda \approx c\tau^{-1}_e$ set by the uncertainty principle \citep[cf.][]{malvimat2014intensity}.

\begin{figure}
    \centering
    \includegraphics[width=0.9\columnwidth]{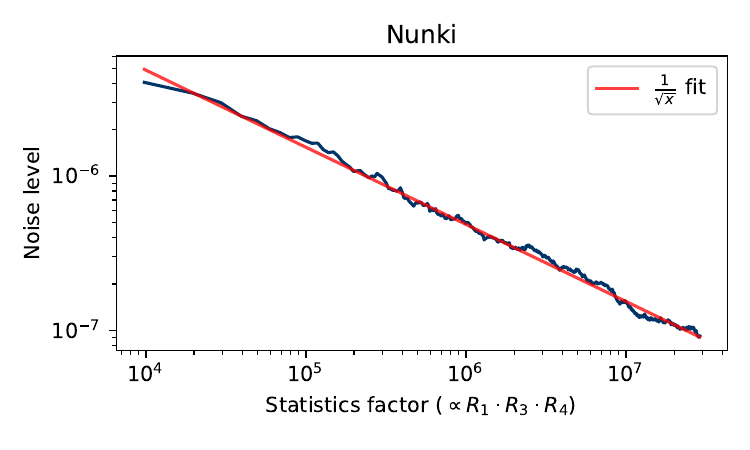}
    \includegraphics[width=0.9\columnwidth]{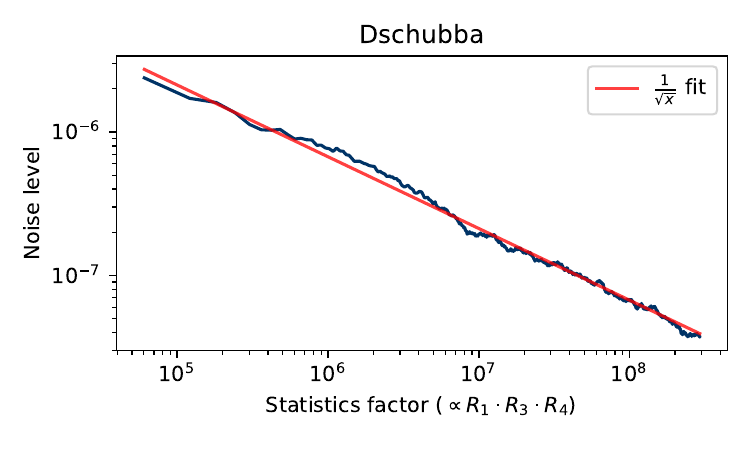}
    \caption{Cumulative noise ($g^{(3)}$ RMS) of both star measurements. The horizontal axis is the product of the mean values of the digitised waveforms and therefore proportional to the rate triple-product.}
    \label{fig:noise}
\end{figure}

An option to reach higher sensitivity is to acquire more measurement data. This is always a valid option as long as the interferometer is shot-noise limited rather than limited by any electronic or other systematic noise. We investigated the noise behaviour for the H.E.S.S.\ measurements, and found the measurements to be shot noise limited over the entire observation time, as the plots in Fig. \ref{fig:noise} indicate: here the RMS of the $g^{(3)}$ measurement in the off-signal region is plotted versus cumulatively more measurement data. As the photon rates were not constant in the measurements, the horizontal axis is not directly $T$, but corrected for the fluctuating rates.

The data follow well the $\sqrt{x}^{-1}$ behaviour, expected from poissonian counting statistics, as the applied fits indicate. However, in order to reach the requested sensitivity on $\cos \phi$, the amount of acquired measurement time needs to be multiplied by a factor of $2300^2$ and $1860^2$ assuming similar photon rates, resulting in total observation times of $1056\,$years for Nunki, and $2447\,$years for Dschubba.

Promising options to improve in S/N, thereby mitigating such long observations times, is the choice of bigger (or more) telescopes, a reduced optical bandwidth, improved time resolution, and spectral multiplexing. As an example for the near future, utilizing the Large-Sized-Telescopes of either CTAO-North or CTAO-South together with very ambitious measurement parameters result in a major reduction in measurement time down to 2 -- 5 months, as summarised in Table \ref{table:signal-to-noise}. 

\begin{table}
\centering
\caption{Comparison of different telescope designs for the examination for closure phase measurements of Nunki and Dschubba.}
\begin{threeparttable}
\label{table:signal-to-noise}
\begin{tabular}{ c  c  c } 
 \hline
Parameter & H.E.S.S. & CTA LSTs\\ [0.1ex] 
 \hline
 Total light collection area  &  $3 \cdot 100\,$m$^2$ &  $4 \cdot 400\,$m$^2$ \\
 $\tau_e$                  &  $5\,$ns              &  $0.1\,$ns \\
 $\Delta\lambda$           &  $10\,$nm             &  $0.1\,$nm \\
 $N_\textnormal{spectral}$ &  $1$\tnote{1}         &  $1000$ \\
 \hline
 $T_{\Delta\cos \phi(\textnormal{Nunki}) \leq 0.1 }$ & $1056\,$years & $2.08\,$months \\
 $T_{\Delta\cos \phi(\textnormal{Dschubba}) \leq 0.1 }$ & $2447\,$years & $4.8\,$months \\
 \hline
\end{tabular}
\begin{tablenotes}
\item[1]As mentioned before, H.E.S.S.\ measured in two colors simultaneously during these measurements, but the wavelength band at $375\,$nm was not considered for the evaluation of $g^{(3)}$.
\end{tablenotes}
\end{threeparttable}
\end{table}

While the assumed time resolution is a rational estimation given the low time dispersion of the parabolic mirror design \citep{karl2022comparing}, the achievable numbers of spectral channels and narrow optical bandwidths, which depend on the optical quality of the telescopes, may be very, potentially too optimistic. Nevertheless, it shows how technical and optical improvements in near-future telescope arrays can reduce the necessary measurement time dramatically. However, since a frequent argument in favor of intensity interferometry is the almost instant image forming capability with large arrays, even higher sensitivities on closure phase measurements are desirable.

The CTAO LST calculation still ignores the additional benefit of a four-telescope interferometer compared to a three-telescope interferometer, providing 4 simultaneous closure phase triangles instead of one. In a dedicated intensity interferometer, such as the proposed SIITAR array with more than $2000$ telescopes \citep{carlile2024purpose}, the ever larger amount of individual closure phase triangles may become a relevant factor. However, since a discussion of optimised telescope array layouts and quantitative considerations are not the scope of this article, we leave this exercise for theorists.

\section{Conclusions}

In this paper we have presented the first third-order correlation measurements of astrophysical targets in intensity interferometry. Even though the signal-to-noise is too low to extract valuable information about the closure phase, which is the seeked improvement to second-order correlations, we aim to intensify discussions about closure phases in the community with this work. We have introduced an analysis algorithm able to extract $\cos \phi$ from the $g^{(3)}$ data, as demonstrated with the laboratory experiments. We believe it is worth performing these kind of laboratory experiments, expanding the parameters to more complex source geometries. This has been done in the past in the laboratory in Lund \citep{dravins2014stellar}, but yet without the investigation of third-order correlations. Such measurements will show the experimental exploitation and limits of $\cos \phi$ in a real world application.

With the current generation of intensity interferometers, measurements of closure phases are highly unreasonable, and even with near future telescopes arrays, such as the CTAO, the value of closure phase data is limited to very long exposures. Rather it seems that in a future dedicated intensity interferometry telescope array, closure phase measurements may add value to the two-photon correlations, that are by default observed.

\section*{Acknowledgements}
We thank Joachim von Zanthier, Verena Leopold and Sebastian Karl for helpful discussions and support with the laboratory measurements. We acknowledge the financial support of the French National Research Agency (project I2C, ANR-20-CE31-0003), the European project IC4Stars (ERC Advanced Grant No. 101140677), and the Deutsche Forschungsgemeinschaft (‘Optical intensity interferometry with the H.E.S.S. gamma-ray telescopes’ –FU 1093/3-1).

\section*{Data Availability}
The data underlying this article will be shared on reasonable request to the corresponding author. Correlation histograms are available in time-intervals of $3.436\,$s. Due to the large size of the digitised waveforms in excess of several TB, the raw data cannot be made available online.



\bibliographystyle{mnras}
\bibliography{citations} 








\bsp	
\label{lastpage}
\end{document}